\documentclass[twocolumn]{aastex63}

\received{ }
\revised{ }
\accepted{\today}
\submitjournal{ApJ}

\shorttitle{Erupting minifilament spectroscopy}
\shortauthors{Kontogiannis et al.}

\begin{document}

\title{High-resolution spectroscopy of an erupting minifilament and its impact on the nearby chromosphere}

\correspondingauthor{Ioannis Kontogiannis}
\email{ikontogiannis@aip.de}

\author{I. Kontogiannis}
\affiliation{Leibniz-Institut f\"{u}r Astrophysik Potsdam (AIP)
An der Sternwarte 16, 14482 Potsdam, Germany}

\author{E. Dineva}
\affiliation{Leibniz-Institut f\"{u}r Astrophysik Potsdam (AIP)
An der Sternwarte 16, 14482 Potsdam, Germany}
\affiliation{Universit\"at Potsdam, Institut f\"ur Physik und Astronomie,
Karl-Liebknecht-Stra\ss{}e 24/25, 14476 Potsdam, Germany
}

\author{A. Diercke}
\affiliation{Leibniz-Institut f\"{u}r Astrophysik Potsdam (AIP)
An der Sternwarte 16, 14482 Potsdam, Germany}
\affiliation{Universit\"at Potsdam, Institut f\"ur Physik und Astronomie,
Karl-Liebknecht-Stra\ss{}e 24/25, 14476 Potsdam, Germany
}

\author{M. Verma}
\affiliation{Leibniz-Institut f\"{u}r Astrophysik Potsdam (AIP)
An der Sternwarte 16, 14482 Potsdam, Germany}

\author{C. kuckein}
\affiliation{Leibniz-Institut f\"{u}r Astrophysik Potsdam (AIP)
An der Sternwarte 16, 14482 Potsdam, Germany}

\author{H. Balthasar}
\affiliation{Leibniz-Institut f\"{u}r Astrophysik Potsdam (AIP)
An der Sternwarte 16, 14482 Potsdam, Germany}

\author{C. Denker}
\affiliation{Leibniz-Institut f\"{u}r Astrophysik Potsdam (AIP)
An der Sternwarte 16, 14482 Potsdam, Germany}





\begin{abstract}

We study the evolution of a mini-filament eruption in a quiet region at the center of the solar disk and its impact on the ambient atmosphere. We used high-spectral resolution imaging spectroscopy in H$\alpha$ acquired by the echelle spectrograph of the Vacuum Tower Telescope (VTT), Tenerife, Spain, photospheric magnetic field observations from the Helioseismic Magnetic Imager (HMI), and UV/EUV imaging from the Atmospheric Imaging Assembly (AIA) of the Solar Dynamics Observatory (SDO). The H$\alpha$ line profiles were noise-stripped using Principal Component Analysis (PCA) and then inverted to produce physical and cloud model parameter maps. The minifilament formed between small-scale, opposite-polarity magnetic features through a series of small reconnection events and it erupted within an hour after its appearance in H$\alpha$. Its development and eruption exhibited similarities with large-scale erupting filaments, indicating the action of common mechanisms. Its eruption took place in two phases, namely a slow rise and a fast expansion, and it produced a coronal dimming, before the minifilament disappeared. During its eruption we detected a complicated velocity pattern, indicative of a twisted, thread-like structure. Part of its material returned to the chromosphere producing observable effects on nearby low-lying magnetic structures. Cloud model analysis showed that the minifilament was initially similar to other chromospheric fine structures, in terms of optical depth, source function and Doppler width, but it resembled a large-scale filament on its course to eruption. High spectral resolution observations of the chromosphere can provide a wealth of information regarding the dynamics and properties of minifilaments and their interactions with the surrounding atmosphere.

\end{abstract}



\section{Introduction} \label{sec:intro}

Filaments are elongated structures seen in absorption in strong chromospheric lines \citep[see][]{2010SSRv..151..333M,parenti14}. These structures contain cool and dense plasma, suspended up to coronal heights by magnetic forces. Although outside active regions the magnetic concentrations are considerably weaker, they can support filaments over a large range of sizes, from giant filaments, which span across hundreds of megameters \citep[e.g., ][]{yazev88,kuckein16,diercke18}, to more compact active-region filaments \citep[see e.g.,][]{kuckein12a}, and small-scale versions called miniature filaments or, more commonly, minifilaments \citep{hermans_martin86, wang2000}. 

\citet{hermans_martin86} were the first to draw attention to these small-scale, elongated features, which were abundant in quiet-Sun H$\alpha$ time-lapse films. They detected sixty-three events in thirty-two days of observations taken from the Big Bear Solar Observatory (BBSO) and presented a detailed account of their characteristics. Their average length was 15\arcsec, their average lifetime 70\,min and it was estimated that 600 of them appear on the Sun every day. Their evolution exhibited formation, a darkening phase, and an eruptive phase whereby they underwent lateral displacement and outward expansion until eventually disappearing. This early study also indicated that most of these small-scale eruptions were associated with sites of magnetic flux cancellation and pointed out analogies in evolution with their large-scale counterparts. \citet{wang2000} performed a similar study but this time also employing magnetograms acquired by the Michelson Doppler Image \citep[MDI,][]{mdi} onboard the Solar and Heliospheric Observatory \citep[SoHO,][]{soho}. They introduced the term ``minifilament'', reported lengths and lifetimes around 19\,Mm and 50\,min, respectively, and emphasized that their loop-like morphology distinguished them to other quiet-Sun structures such as macrospicules. 

The advent of space-borne UV and soft X-ray observatories facilitated multi-wavelength observations of minifilaments, and the study of their evolution in chromosphere and in the overlying corona. In \citet{sakajiri04} and \citet{ren08}, H$\alpha$ filtergrams were combined with EUV imaging, showcasing the sequence of events during the formation and eruption of minifilaments, namely, flux cancellation at the photosphere, darkening and expansion of the chromospheric absorption features, rotating motions, and radial eruption of the whole or part of the structures. During this motion, flare-like brightenings and EUV/soft X-ray dimmings also occurred. 

Subsequent studies demonstrated the close association between minifilaments and other coronal eruptive events. Roughly a quarter of the quiet-Sun minifilaments were involved in eruptions which produced mini-CMEs and even small-scale coronal waves  \citep{innes09,Podladchikova2010,Schrijver2010}. \citet{moore10}, although not discussing minifilaments per se, estimated that one-third of the coronal jets were produced by the eruption of sheared-core magnetic arcades, similarly to active regions. \citet{raouafi10} supported that micro-sigmoids are pro-genitors of coronal jets, which can explain the helical structure of the ejecta \citep{patsourakos08} and the presence of mini-CMEs. Analysis of two-viewpoint observations of a minifilament eruption by \citet{hong11} supported that mini-CMEs and blowout jets may share a common origin, i.e., a core magnetic structure. \citet{hong14} found that brightness enhancements of coronal bright points were related to minifilament eruptions, which were attributed to reconnection due to flux convergence and cancellation below the loop structure. \citet{sterling15} presented more evidence on the role of minifilament eruptions in coronal ejecta and heating, while recent results indicate that more than two thirds of the coronal jets in the quiet Sun and coronal holes were associated with minifilament eruptions \citep{mcglasson19,kumar19a}. 

Another interesting aspect is the interaction of minifilaments with the ambient atmosphere and the effect of their eruptions on nearby solar structures. Earlier studies suggested that part of the erupting minifilament material returns to the chromosphere \citep{hermans_martin86,wang2000,sakajiri04}. More recent studies based on high-resolution EUV imaging showed aspects of this interaction in the corona. When formed in the vicinity of active regions, minifilaments can interact with the large-scale magnetic field and reconnect with the large-scale magnetic field of active regions \citep{chen19}, or other filaments producing jets \citep{yang_b19,sterling19}. \citet{zheng13} found that the eruption of a micro-sigmoid led to the propagation of a coronal wave, which interacted with nearby loops, producing downflows. \citet{yang_l19} demonstrated how a minifilament interacted with large-scale magnetic loops of an active region, causing heating and a blow-out jet. They found that the twist of the minifilament was partly transferred to the newly formed loops. 

Based on the aforementioned studies and results, it is largely accepted that the formation and eruption of minifilaments is dictated by mechanisms common to the ones found in active-region filaments. These involve magnetic footpoint motions and shear \citep{kumar18}, converging motions during flux emergence and flux cancellation \citep{denker09,hong11,hong14,panesar17,mcglasson19,sterling19}. As such, the formation (i.e., bodily emergence of a twisted structure versus gradual formation) and the eruption of minifilaments (i.e., which mechanisms drive the instability) are subject to the same uncertainties and ambiguities as active-region filament eruptions (which can lead to flares and CMEs). Studying these structures offers the opportunity to study the same fundamental physical processes in smaller scales, rendering them an ideal target for meter-class (and beyond) telescopes, whose FOV is limited.

\begin{figure*}
\centering
  \includegraphics[width=19cm]{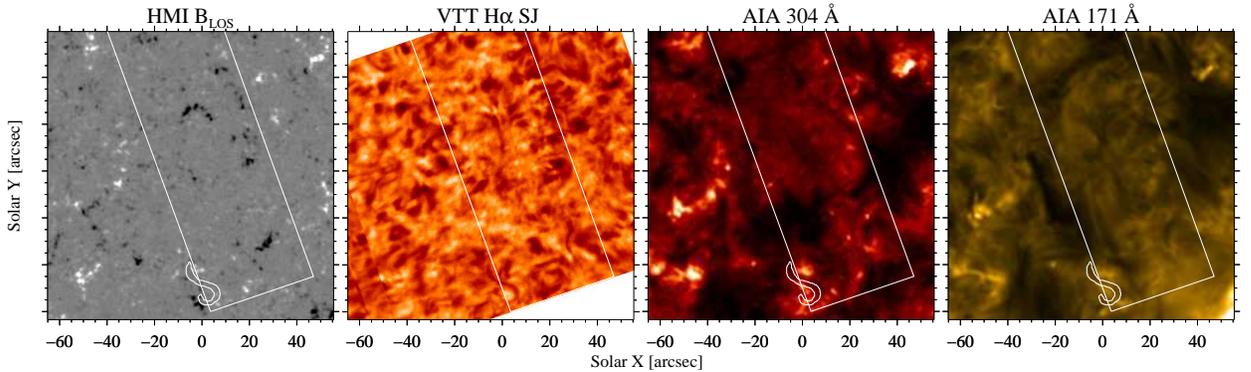}
  \caption
    {The solar region near the disk center, observed on May 26, 2019 at 10:15\,UT. From left to right: HMI magnetogram (scaled between $\pm$100\,G), the VTT slitjaw H$\alpha$ image and the AIA filtergrams in $\lambda$304 and $\lambda$171\,{\AA}. The overploted rectangle indicates the scanning region of the echelle spectrograph while the contours in the HMI magnetogram and AIA filtergrams outline the minifilament, as seen in the H$\alpha$ slitjaw images. The images are aligned to the solar North.}
    {\label{fig:fov}}
\end{figure*}

During the past decade the study of minifilaments was mainly limited to high-quality observations of the Solar Dynamics Observatory \citep[SDO,][]{sdo}, complemented by context H$\alpha$ imaging. Therefore, the dynamic evolution of erupting minifilaments and the response of the ambient chromosphere has not been studied in detail yet. Here we present such a detailed study, using time-series of high spectral-resolution imaging in H$\alpha$, which were taken using a new observing setup in the Vacuum Tower Telescope, Tenerife, Spain \citep[VTT,][]{vtt}. Spectrally resolved H$\alpha$ observations of minifilaments are very few; those that exist are mostly context observations and, to our knowledge, this is the first time such a study is performed, showcasing the intriguing nature of minifilaments and the potential of high-spectral resolution observations. 

The paper is organized as follows. In Section~\ref{sec:obs_anal}, we describe the observing setup and the data reduction steps. In Section~\ref{sec:results}, we describe the evolution of the minifilament eruption and its impact on the nearby chromosphere, based on reconstructed maps of physical quantities, spectral properties, and cloud model parameters. These results are then discussed in the context of the existing literature in Section~\ref{sec:conclusions}, where we also highlight our conclusions. 



\section{Observations and analysis}\label{sec:obs_anal}

The observations were obtained between 07:00 and 13:00\,UT on May 26, 2019, from the VTT, in Tenerife, Spain. They were part of a coordinated observing campaign, which included the VTT, and the Interface Region Imaging Spectrometer \citep[IRIS,][not used in this study]{iris}. The target of the campaign was a quiet-Sun region at the solar disk center. The excellent seeing conditions enabled the Kiepenheuer Institute Adaptive Optics System \citep[KAOS,][]{kaos} to lock on quiet-Sun granulation.

Three pco.4000 CCD cameras were mounted at the echelle spectrograph to simultaneously record spectra in three spectral regions, namely the H$\alpha$ ($\lambda$6562.8\,{\AA}), H$\beta$ ($\lambda$4861\,{\AA}) and the magnetic sensitive Cr\,{\sc i} line ($\lambda5781$\,{\AA}). Broadband interference filters were used in front of each camera to block light from overlapping spectral orders. The binning of the cameras was $\times$4 in the spatial and $\times$8 in the spectral dimension. The slit width of the echelle spectrograph was 80\,$\mu$m, resulting in a scanning step equal to 0.36\arcsec\ on the solar surface. This was also the pixel size along the slit. The spectral resolution was 15\,m{\AA}\,pixel$^{-1}$ This setup offers the opportunity to obtain relatively fast time series of scans, from a few minutes down to a few tens of seconds, depending on the field-of-view (FOV), with very high spectral resolution. In this study we focused on the H$\alpha$ time series of scans; the exposure time for each step of the scan was 70\,ms facilitating 50\arcsec$\times$216\arcsec\ consecutive scans with a 20-second cadence. 

All steps of VTT reduction and analysis of the echelle spectra, namely dark and flat field correction, removal of the spectrograph tilt, wavelength and intensity calibration, were performed in Interactive Data Language (IDL) routines, which are part of the sTools software library \citep{stools}. The data reduction process includes Principal Component Analysis (PCA) as a means to clean spectra from noise and blends with telluric lines, to facilitate further spectral analysis. After the process, the clean spectra were centered on the nominal H$\alpha$ line center at 6562.8\,{\AA}, extending $\pm$2.5\,{\AA} into the wings, and were then used to produce maps of quantities such as Doppler velocity, line-core intensity, full width at half maximum (FWHM), equivalent width, bisector velocities, etc. For each scan a noise-stripped, average quiet-Sun profile, $I_{QS}$ was calculated, and from all observed profiles, $I_{obs}$, contrast profiles, $(I_{obs}-I_{QS})/I_{QS}$, were constructed. From these, radiative transfer and line formation parameters were derived by means of cloud model \citep[CM,][]{beckers64} inversions. We will discuss the cloud model and its derived parameters in more detail in Section~\ref{sec:cloud}. Further details on the reduction pipeline of this type of VTT observations and the use of PCA to denoise spectra and derive CM parameters is given in \citet{dineva20}. 

Context UV and EUV observations were provided by the Atmosphering Imaging Assembly \citep[AIA,][]{aia} of SDO. In this study we used time series of images recorded in the 1600\,{\AA}, 304\,{\AA}, 171\,{\AA} and 193\,{\AA} channels. The cadence of the observations was 24\,s for the 1600\,{\AA} channel and 12\,s for the rest. Furthermore, we used the LOS magnetograms provided every 45\,s by the Helioseismic and Magnetic Imager \citep[HMI,][]{scherrer12,schou12}. 

The region encompassing the minifilament eruption was scanned by the echelle spectrograph between 10:03\,UT and 10:50\,UT (Figure~\ref{fig:fov}), thus capturing the eruption which took place around 10:25\,UT. However, the evolution of the region before the formation and eruption of the minifilament was monitored using AIA and HMI as well as slit-jaw (SJ) images of the echelle spectrograph in H$\alpha$ and Ca\,{\sc ii}\,K. 

The co-alignment of the observations was performed first by comparing the slit-jaw images of the echelle spectrograph in Ca\,{\sc ii}\,K with the AIA filtegrams at 1600\,{\AA} and, therefore, the rest of the AIA channels. The 1600\,{\AA} filtergrams were then co-aligned with the HMI magnetograms and the IRIS SJ images. The latter were not used in the present study because the minifilament was located just outside the southern edge of the IRIS FOV.

\begin{figure*}
\centering
  \includegraphics[width=18cm]{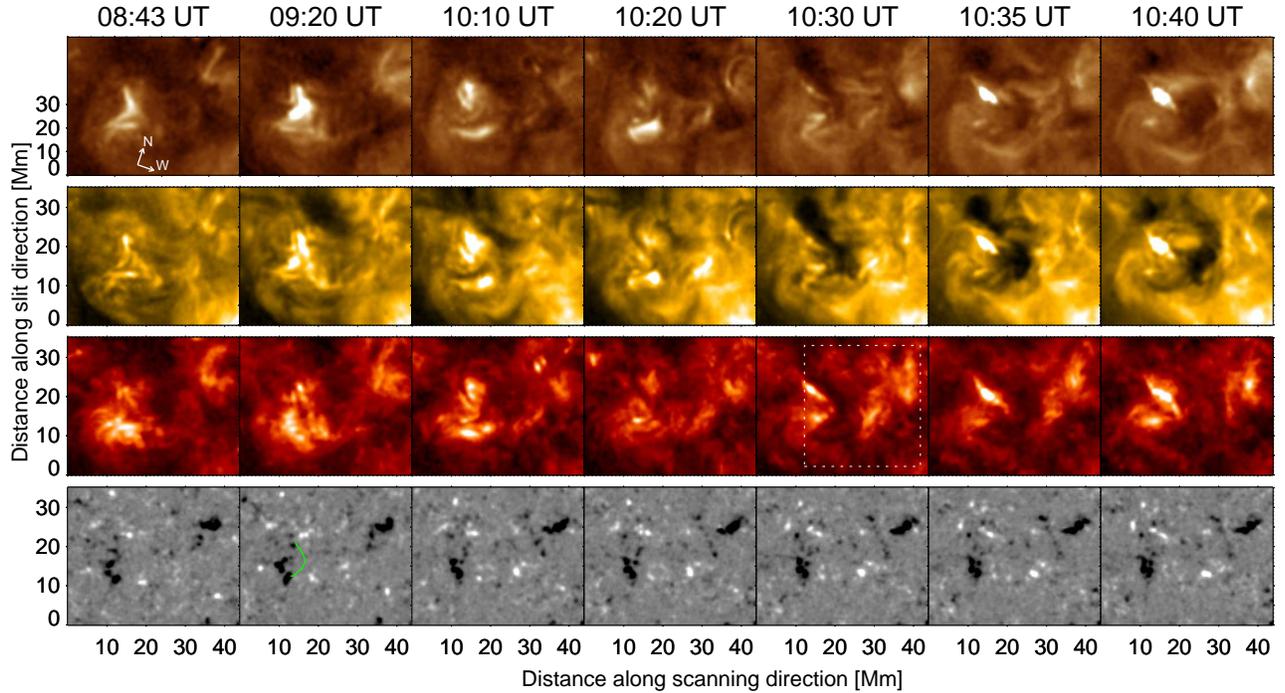}
  \caption
    {From top to bottom: context EUV images in 193\,{\AA}, 171\,{\AA}, 304\,{\AA} and photospheric magnetograms of the minifilament and its surroundings. The magnetograms are scaled between $\pm$50\,G to enhance the visibility of the quiet-Sun magnetic concentrations. The green curve in the photospheric magnetogram (10:20\,UT) indicates roughly the position of the minifilament. All cut-outs are aligned to the reference system defined by the slit orientation and scan direction of the echelle spectrograph. The dashed rectangle (third row) represents part of the region that was scanned by the echelle spectrograph and is shown in Figure~\ref{fig:ha_evo}. The Cartesian axes at the top left indicate the orientation of the solar North. }
    {\label{fig:uv_evo}}
\end{figure*}

\section{Results}\label{sec:results}

\subsection{Overview of the minifilament formation and eruption}\label{sec:overview}

Figure~\ref{fig:fov} shows the quiet region near solar disk center as observed on May 26, 2019 by VTT and SDO. The minifilament was formed in the region between a slightly more extended negative magnetic-polarity footpoint and some smaller positive ones. When fully formed, it had an S-shape, with the southern, convex part being more extended than the northern, concave one. The H$\alpha$ SJ images show the typical mottled appearance of the chromosphere, wherein the minifilament resides with a contrast similar to that of the other fine structures (mottles or fibrils). In 304\,{\AA}, the region showed the typical network emission, corresponding to small coronal bright points seen in the 171\,{\AA} channel.

An overview of the evolution of the region prior and during the eruption is presented in Figure~\ref{fig:uv_evo}. The photospheric magnetic field (bottom row) did not show any pronounced variation, e.g., flux emergence or cancellation, although it is possible that these mechanisms acted on a scale near or below the resolution and sensitivity limit of HMI, as the small-scale magnetic field was reconfiguring to produce the minifilament. Applying the Differential Affine Velocity Estimator \citep[DAVE,][]{schuck06} method we found persistent converging motions of the two magnetic polarities towards the neutral line (found in the region marked by the green curve in Figure~\ref{fig:uv_evo}, 09:20\,UT, bottom row). On top of this motion, the individual points exhibited apparent proper motions in smaller scale, rotating and shuffling. By the end of the observations, the negative polarity became more compact and the opposite polarity footpoints approached each other. Converging motions of the footpoints, as the ones found here, are often considered to be the cause of minifilament eruptions \citep{hong11,hong14,kumar18}.

\begin{figure*}
\centering
  \includegraphics[width=19cm]{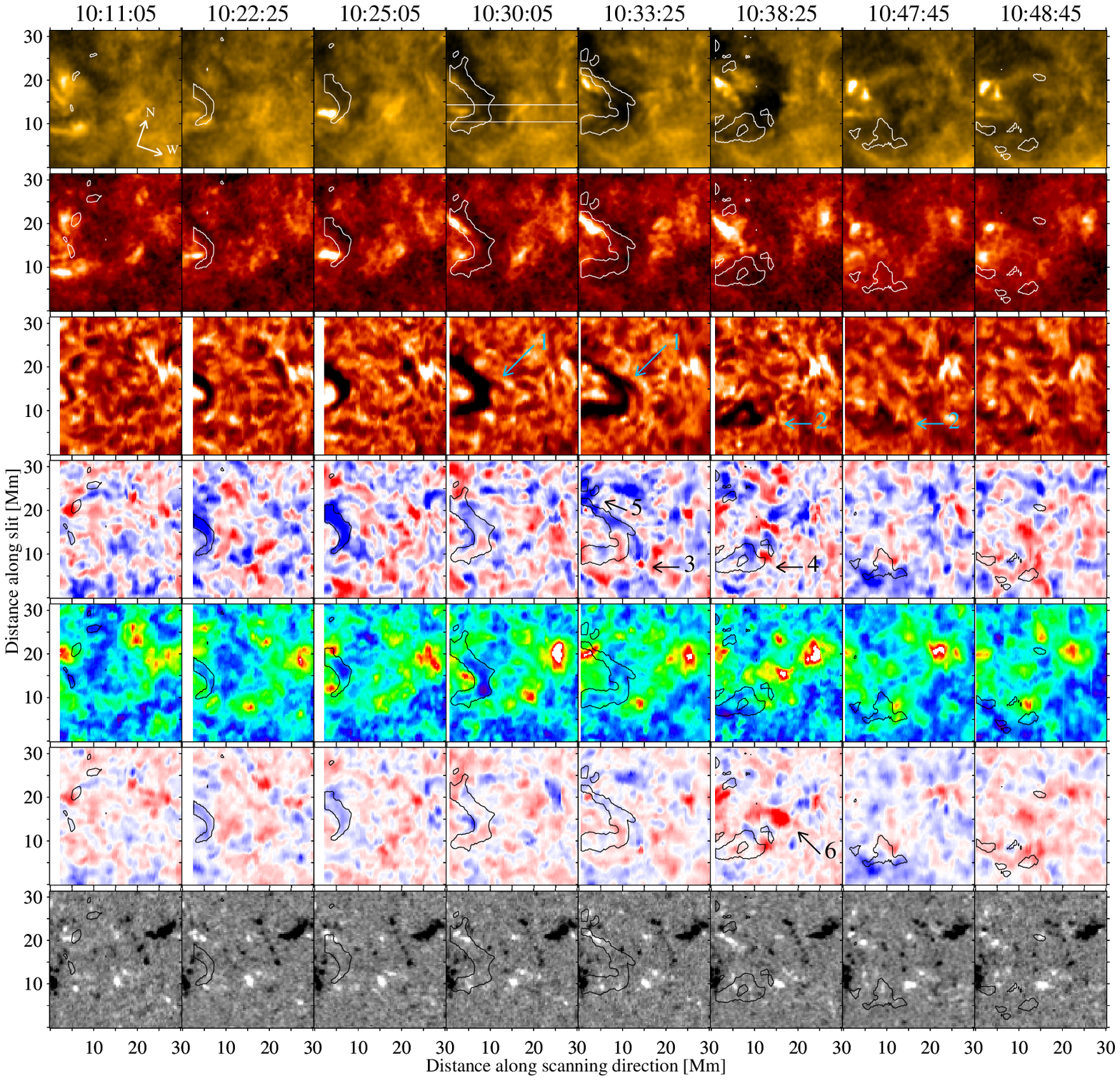}
\caption
    {Overview of the minifilament eruption. From top to bottom are the AIA filtergrams in 171\,{\AA} and 304\,{\AA} (for context), the  H$\alpha$ Doppler-shift-compensated core intensity, Doppler velocity (scaled between $\pm$5\,km\,s$^{-1}$), FWHM (scaled between 1.3 (blue) and 1.8\,{\AA}(red)), bisector velocity at the half-maximum (BVHM, scaled between $\pm$5\,km\,s$^{-1}$) and HMI photospheric magnetograms. To facilitate comparison, the cut-outs are aligned to the reference system defined by the slit orientation and scan direction of the echelle spectrograph. Numbers and arrows denote features of interest (see text). The Cartesian axes at the top left indicates the orientation of the solar North and West directions.}
    {\label{fig:ha_evo}}
\end{figure*}

The EUV emission of the region exhibited intense variability. In the hotter emission channels (171\,{\AA} and 193\,{\AA}) a structure resembling a micro-sigmoid was visible at the beginning of the observations. This structure underwent changes in shape and brightness, giving the impression of an interaction between the bright loops that were initially connecting the magnetic footpoints of the region (08:43\,UT). The minifilament was gradually forming and eventually seen at 09:20\,UT in the 171\,{\AA} and 304\,{\AA} channels, embedded in a bright envelop. At the same time, the minifilament appeared clearly in the H$\alpha$ SJ images (not shown). The brightenings seen above and around the structure in EUV can be attributed to successive reconnection events. These can build up the magnetic flux and provide the minifilament with plasma, reconfiguring the overlying magnetic field to facilitate its eruption, similarly to what is observed in active regions. 
The entire minifilament is clearly discernible in EUV images after 10:10\,UT, first in the 304\,{\AA} and then in the 171\,{\AA} and 193\,{\AA}. After the last brightening seen at the lower part of the minifilament (10:20\,UT), it eventually erupted. This sequence of events supports a scenario where, upon its formation, the structure started to rise slowly, reconnected with the overlying magnetic field, and gradually made its way to eruption.   

As seen clearly in the 193\,{\AA} and 171\,{\AA} filtergrams (Figure~\ref{fig:uv_evo}, first and second row), the minifilament did not erupt homogeneously. The northern part erupted first, became more opaque than the southern part and it traversed a much smaller distance, exhibiting a quasi-rotational motion in the east-to-west direction. Conversely, the southern part, which was more clearly visible in the chromosphere, had a conspicuous semi-circular shape and its expansion and eruption had a more pronounced horizontal component, sweeping the ambient atmosphere and producing more extended dimmings in all EUV channels. Based on these observations, we conjecture that the northern part of the minifilament had a significant propagation component along the LOS while the southern part had also a predominant horizontal expansion component. The H$\alpha$ spectroscopic observations described in the following section will corroborate the latter conjecture.  

Given the time of its first unambiguous detection in H$\alpha$ and the time of its eruption, it is estimated that the lifetime of the minifilament was of the order of one hour, similarly to what was reported by \citet{wang2000} and \citet{hermans_martin86}. However the entire formation process could not be monitored by the H$\alpha$ spectroscopy because these observations started at 10:03\,UT, when the minifilament was already formed. Instead, we were able to monitor the later stages of evolution towards the eruption of the southern, convex part, which we will describe in detail in the next section.


\subsection{Slit-reconstructed maps of H$\alpha$ spectral characteristics}\label{sec:physmaps}

The FOV of the echelle spectrograph contained the southern part of the minifilament and its eruption. In the following we will be referring to this part as the ``minifilament'', for brevity. In Figure~\ref{fig:ha_evo} we present the slit-reconstructed maps of some key spectral parameters of the H$\alpha$ line, namely, the Doppler shift-compensated minimum intensity (simply: core intensity), the Doppler velocity (derived through a Gaussian fit), the FWHM and the bisector velocity measured at the half-maximum of the profile (BVHM). Although the line formation processes of H$\alpha$ are complicated, it can be considered that this set of parameters represent different type of information encapsulated in the H$\alpha$ profiles. The core intensity and Doppler velocity characterize the highest chromospheric layers of line-formation, while the FWHM is associated with heating and the bisector velocity BVHM characterizes motions, both seen at lower chromospheric heights \citep{cauzzi09,leenaarts12}. We study these maps along with the images in 304 and 171\,{\AA}, to get a closer and more detailed view of its evolution across all atmospheric layers. 

As seen in the two top rows of Figure~\ref{fig:ha_evo}, this part of the minifilament became darker, thicker, and it started expanding radially, after 10:11\,UT. Nine minutes later it erupted almost radially, producing a dimming that swept across the nearby atmosphere. This dimming and its propagation was more pronounced in the coronal 171\,{\AA} emission, extending out to 20\,--\,25\,Mm from its origin, and less in the upper chromospheric 304\,{\AA} emission. Behind the dark rim of the propagating dimming, coronal and chromospheric brightenings appeared, near both ``footpoints'' of the minifilament. These flare-like EUV brightenings are common in minifilament eruptions \citep[see e.g.][]{hong11} and can be attributed to the reconnection with the magnetic field of the source region.  

The third row of Figure~\ref{fig:ha_evo} shows the evolution of the H$\alpha$ line core intensity. In these maps we see more clearly the chromospheric counterpart of the thickening, darkening and eruption. Brightenings peek out behind the structure, corresponding to the ones seen in the hotter channels of AIA. At 10:30\,UT an arc-like structure is protruding from the main body of the minifilament (``1''), which was fully formed three minutes later as a second arc-shaped absorption structure ahead of the main body of the minifilament. Five minutes later (10:38\,UT) the chromospheric counterpart of the minifilament was already fragmented, leaving behind only its lower arm (``2''). This part exhibited a twisted structure and eventually moved laterally, its fragments dissolving into the chromospheric H$\alpha$ background. Both the protruding arc and the dissolving arm support that the minifilament consists of finer threads.

The Doppler velocity maps in Figure~\ref{fig:ha_evo} (fourth row) show that initially the minifilament exhibits a bulk upward motion with velocity in excess of 5\,km\,s$^{-1}$. After this initial bulk motion, the southern part of the structure moves downward (see e.g. 10:30\,UT) while the northern part maintains its upward motion. The secondary arc-like structure that developed after 10:30\,UT continued the upward motion until its disappearance from the H$\alpha$ maps. Interestingly, at the lower part of this secondary, blue-shifted arc strong downflows are found (10:33\,UT ``3''), which indicate interaction with the neighbouring small-scale magnetic structures of the quiet-Sun. The spiral-shaped arm that remained from the eruption exhibited strong alternating upward and downward motions (10:38\,UT, ``4''), further supporting a spiral structure of the minifilament. Another interesting feature clearly seen in the Doppler velocity maps is a chromospheric jet-like structure at the upper part of the minifilament (10:33\,UT, ``5''). At the base of this structure, a strong downflow ``island'' is embedded in the bulk upward motion and this region coincides with the intense brightenings seen in 171 and 304\,{\AA}. The Doppler velocity maps at 10:47\,UT and 10:48\,UT indicate that after the eruption, strong downflows are located ahead of the erupting minifilament and at its footpoints. We will further discuss these in the following description of the FWHM and BVHM maps.

\begin{figure*}
\centering
  \includegraphics[width=18cm]{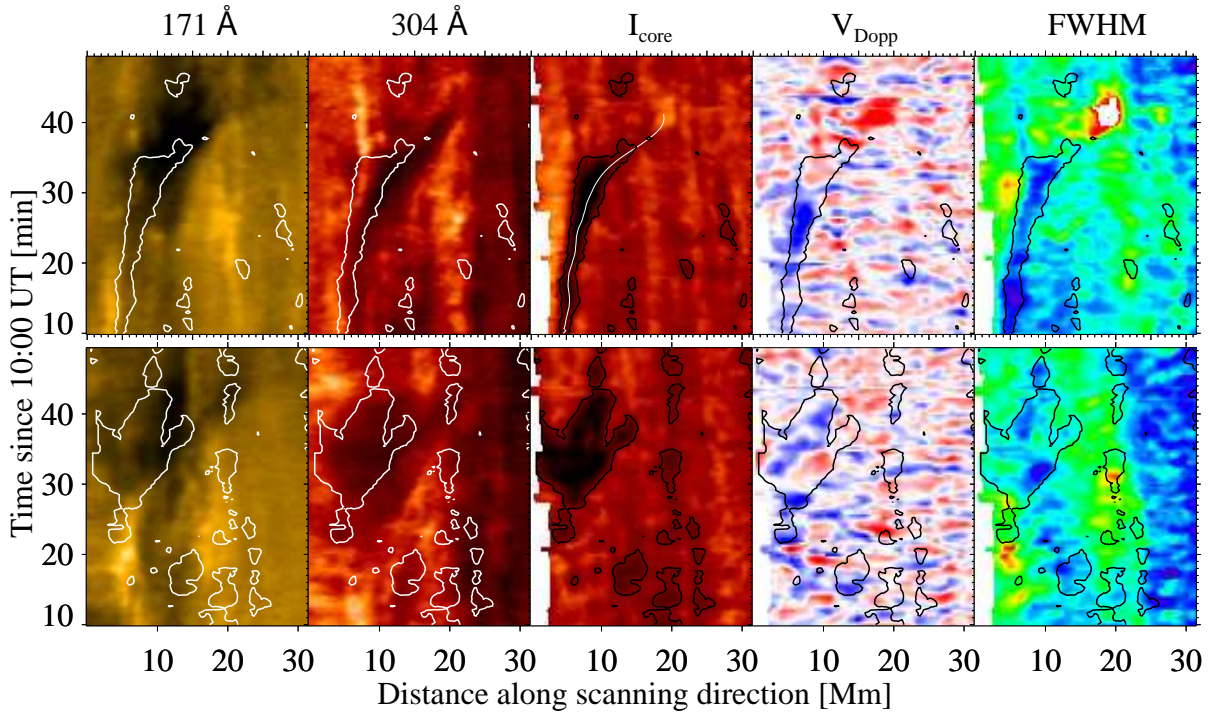}
  \caption
    {Space-time ($X$\,-\,$t$) slices in AIA 171\,{\AA}, 304\,{\AA}, H$\alpha$ core intensity, Doppler Velocity (scaled between $\pm$5\,km\,s$^{-1}$), and FWHM (scaled between 1.3 and 1.8\,{\AA}) across the apex (top row) and the spiral arm (bottom row) of the minifilament (see Figure~\ref{fig:ha_evo}, top row, 10:30\,UT). Contours are used to indicate the $0.155\cdot$ I$_{c}$ of the H$\alpha$ core intensity, which delineates the minifilament. The white line in top row, third panel, indicates the trajectory of the apex.}
    {\label{fig:ha_slice}}
\end{figure*}

The FWHM maps of the region (Figure~\ref{fig:ha_evo}, fifth row) show that the minifilament consisted mostly of plasma with narrower H$\alpha$ absorption profiles. During the expansion, this plasma was found mostly around the apex, while towards the ``footpoints'' (i.e., the region closer to the edge of the FOV and in the vicinity of the photospheric magnetic elements), FWHM values were considerably higher. This was more pronounced at the locations where the flare-like brightenings are seen and also at the base of the jet-like feature at the upper part of the minifilament. A larger FWHM is also detected at the site of interaction between the secondary arc and the nearby quiet-Sun magnetic fields, as well as at locations of strong downflows that followed the eruption. 

The reconstructed maps of the BVHM (Figure~\ref{fig:ha_evo}, sixth row) largely follows the velocity pattern exhibited by the Doppler velocity, but with some notable differences. After the eruption they show very pronounced redshifts, over more extended patches on the FOV (10:38\,UT, ``6''). These are attributed to strong downflows after the eruption and indicate that the corresponding spectral profiles are not only redshifted but also highly asymmetric. Since the BVHM is sensitive to motions lower in the chromosphere \citep{dineva20,verma20}, these strong downflows are attributed to the interaction of the minifilament material with the chromosphere below. In the following, we will discuss these profiles in detail and the regions where they occurred.

\subsection{Space-time slices and high-resolution spectra of the minifilament apex}\label{sec:xtslice}

In Figure~\ref{fig:ha_slice} we plot space-time ($X$\,--\,$t$) slices, to study the motion of the apex and the lower spiral arm of the minifilament. These two-arcsec wide slices were taken along the horizontal lines seen in Figure~\ref{fig:ha_evo} (top row, 10:30\,UT). Overall, the traces of the minifilament in 171\,{\AA} are considerably thicker than the ones in 304\,{\AA} and H$\alpha$, showing the much more extended impact of the eruption to the coronal environment. The shift between the traces of the eruption in H$\alpha$ and the hotter channels is also a result of different formation heights, as the eruption is more extended in the higher layers. The H$\alpha$ Doppler velocity maps show how the initial upward bulk motion is followed by intense downflows at the apex of the minifilament. Although the minifilament then disappeared from the $X$\,-\,$t$ slices of core intensity, patches of large FWHM and intense downflows are seen along the extension of its trajectory. These downflows are likely the result of the ballistic motion of the falling minifilament material. This is the first time that we see such a small-scale interaction between filamentary material and the adjacent chromosphere in such a detail. 

Additionally, the H$\alpha$ slices at the apex show that the evolution of the eruption takes place in two phases, i.e., a slow lateral motion followed by a fast acceleration after 10:30\,UT, which occurred after the extended flare-like brightenings in 304\,{\AA} and the H$\alpha$ line core (see corresponding panels in Figure~\ref{fig:ha_evo}). This two-phase evolution is evident by the curvature of the H$\alpha$ line-core intensity slice, which we determined as follows: For each slice we mark the middle between the boundaries of the 0.155$\cdot$ I$_{c}$ contour. Then, we extend this trace after the end of the contour following the trace of the downflows until we reached the high-FWHM patch. The resulting curve (white solid line in Figure~\ref{fig:ha_slice}) was smoothed using a two-minute running average. We employed a linear fit to estimate the average speed of the apex, and found that the first slow motion has a speed around 2.2\,km\,s$^{-1}$ while the second one corresponds to an average speed of 17.6\,km\,s$^{-1}$. For the EUV channels, the determination of the speed is more challenging due to the thickness of the traces, which can result in varying speed estimates. In 304\,{\AA}, we estimated a projected speed around 23\,km\,s$^{-1}$, which is largely consistent with the one determined for the H$\alpha$ core intensity. In 171\,{\AA}, the projected speed can be even higher, up to about 35\,km\,s$^{-1}$. A higher speed for the parts of the minifilament that appear in the coronal channels is consistent with the appearance of the secondary blue-shifted arc, shown in the slit-reconstructed cut-outs of Figure~\ref{fig:ha_evo} (10:33\,UT). This feature likely represented the part of the minifilament that continued to propagate upwards and subsequently disappeared from the H$\alpha$ spectral region, as its material was heated to higher temperatures. 

The $X$\,--\,$t$ slices of the arm (Figure~\ref{fig:ha_slice}, bottom row) show an overall thicker and more complicated trace. As already mentioned previously, this part exhibited an apparent (un)twisting and lateral motion until eventually fading. The corresponding $X$\,--\,$t$ slices show an alternating upward and downward velocity pattern, supporting this interpretation.

\begin{figure*}
\centering
  \includegraphics[width=17cm]{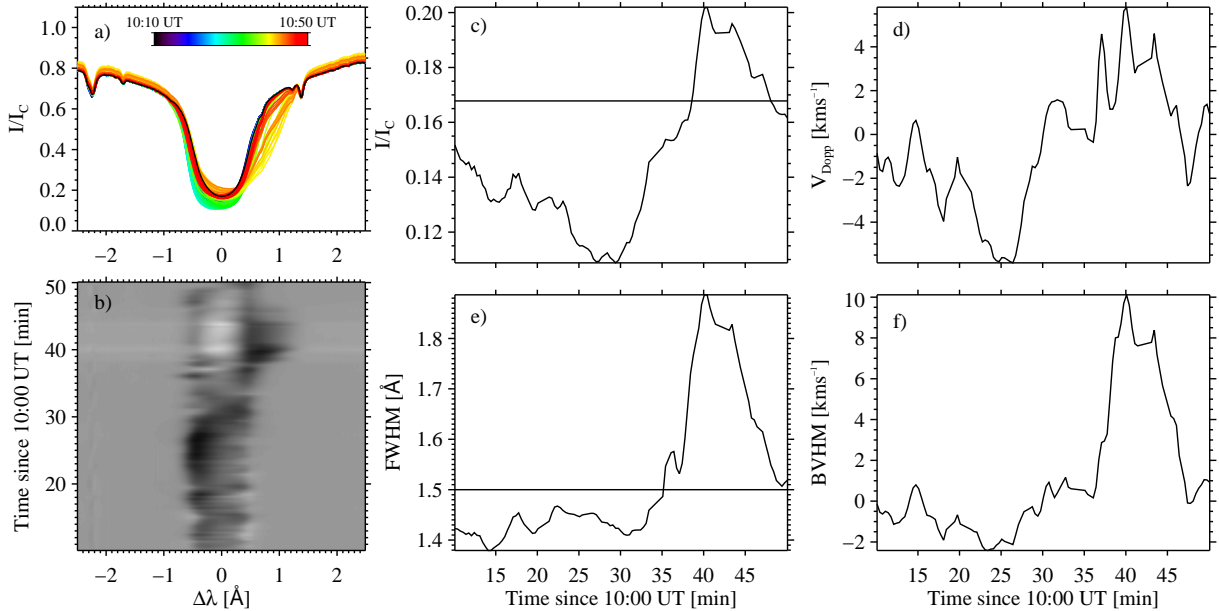}
  \caption
    {The evolution of H$\alpha$ line profiles and spectral characteristics along the trajectory of the minifilament apex. a--b) Evolution of the original spectral line profiles and the PCA-denoised contrast profiles, respectively. c-f) Evolution of the corresponding line-core intensity, Doppler velocity, FWHM and BVHM. The horizontal lines in panels (c) and (e) denote quiet-Sun averages, derived from observations.}
    {\label{fig:ha_prof}}
\end{figure*}

Next, we examine the spectra and spectral characteristics along the trace of the apex, which was determined in Figure~\ref{fig:ha_slice}. Panels (a) and (b) of Figure~\ref{fig:ha_prof} contain two different representations of the evolution of these H$\alpha$ line profiles, one as overplotted profiles, where time is denoted by colors from blue to red, and one as stacked contrast profiles, where time is the vertical axis. In panel (a) we have also included a time-average quiet-Sun profile (plotted in black), taken in a region of the FOV with no pronounced absorption features. The H$\alpha$ profiles (Figure~\ref{fig:ha_prof}a) vary with time, in respect with the quiet-Sun average, first becoming blue-shifted, deeper and narrower and then red-shifted, shallower, wider and clearly asymmetric. The contrast profiles (Figure~\ref{fig:ha_prof}b) show more clearly the progression of the velocity of the opaque minifilament material of the apex. Dark patches, which correspond to increased absorption, shift from the blue wing (10:20--10:30\,UT) to the far red ($\Delta \lambda > 1$\,{\AA}), after the eruption and break-up of the structure. 

\begin{figure}
\centering
  \includegraphics[width=8cm]{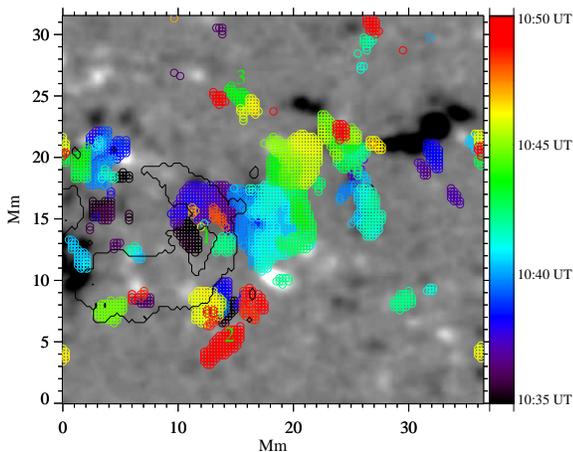}
  \caption
    {Map of the times of maximum BVHM, higher than 3$\sigma$ from the average (i.e., 3\,km\,s$^{-1}$), after 10:35\,UT, overlaid on the corresponding HMI magnetogram (scaled between $\pm$50\,G). The black contour shows the minifilament, as seen in H$\alpha$ core, at the same time to provide context. Colours represent the elapsed time, from black (10:35\,UT) to red (10:50\,UT).}
    {\label{fig:mat_fall}}
\end{figure}

A parametric representation of the temporal variation of the spectral profiles of the apex is given in panels (c) -- (f) of Figure~\ref{fig:ha_prof}. In panels (c) and (e) we have also marked the level of time-averaged quiet-Sun H$\alpha$ core intensity and FWHM, calculated in a region with no pronounced absorption features. The core intensity (Figure~\ref{fig:ha_prof}c) decreases continuously until 10:30\,UT, as the filament darkens and thickens, from 0.15$\cdot$ I$_{back}$ to 0.11$\cdot$ I$_{back}$, which corresponds roughly to a 25\% change in line-core intensity during the darkening phase of the minifilament. After that, the line-core intensity reaches and surpasses the quiet-Sun levels by about 15\%, as the minifilament disappears. This increase resulted from the impact of the minifilament material on the nearby chromosphere. Similarly, the upward Doppler velocity observed clearly until 10:25\,UT, gradually turns downward, eventually leading to strong downflows observed after the eruption (Figure~\ref{fig:ha_prof}d). The temporal variation of the FWHM and BVHM  (Figure~\ref{fig:ha_prof}e,f) is not very pronounced until 10:30\,UT. The FWHM is lower compared to the average quiet-Sun, indicating that the minifilament contains cool chromospheric material, but the FWHM increases rapidly after 10:35\,UT (reaching $\sim$1.9\,{\AA}). The bisector velocity, although it qualitatively follows the evolution of the core velocity, is not very sensitive to macroscopic motions of the minifilament, hence the very low negative values until 10:25\,UT. It is, however, exhibiting a steep rise, reaching as high as 10\,km\,$s^{-1}$, when the minifilament material impacts on the nearby lower chromospheric structures.

\subsection{Interaction with the nearby chromosphere}\label{sec:nearchrom}

The analysis presented so far indicates that a defining characteristic of the impact of the minifilament material on the nearby chromosphere is not the line-core absorption but the bisector velocity at the half-maximum level (BVHM). As already mentioned, this velocity refers to lower chromospheric heights than the Doppler velocity of the line core and is associated with asymmetric profiles. We utilize this to study the spatial distribution and extent of this effect. We used a threshold of 3\,km\,s$^{-1}$, which is the 3$\sigma$ level from the average BVHM (0\,km\,s$^{-1}$), and located the points that exhibited a maximum BVHM higher than this threshold, after 10:35\,UT. 

In Figure~\ref{fig:mat_fall}, these locations are found over the projected expansion of the minifilament, after its eruption (see the black contour provided for context). They are not homogeneously distributed within the FOV; instead they are strongly associated with, or adjacent to, nearby, small-scale magnetic concentrations. Since these are connected via small-scale, low-lying magnetic loops \citep[see e.g.,][]{kontogiannis18}, the observed pattern is the result of the interaction between the returning minifilament material and the chromospheric magnetic loops. The color of the locations in Figure~\ref{fig:mat_fall} ranges from black to red, showing also the ``propagation'' of this effect, from locations closer to the minifilament to sites located farther away. The strongest effect is directly around the apex (``1'') and covers a distance of more than 10\,Mm. The remaining spiraling arm of the minifilament has also an impact, as indicated by the high-BVHM patches at ``2'', seen towards the end of the time-series when the corresponding part of the minifilament dissolves into the chromospheric background. A smaller patch can be associated with the upper arm of the minifilament at ``3''. These findings provide conclusive evidence to the conjecture made by \citet{wang2000}, that part of the material of the minifilament returns to the chromosphere and provide an explanation of the downflows detected in H$\alpha$ line-wing filtergrams by \citet{sakajiri04} and \citet{ren08}.  

\subsection{Cloud model analysis of the minifilament}\label{sec:cloud}

The cloud model can be used to invert spectra from a chromospheric structure, under the simplifying assumption that the structure is suspended in the atmosphere like a cloud of plasma, absorbing the background radiation \citep[see][for a review]{tziotziou07}. The model assumes that the emerging spectral line profile can be fully represented by four parameters, namely the optical depth $\tau$, the source function $S$, the Doppler velocity $\upsilon_{D}$ and the Doppler width $\Delta \lambda_{D}$, which are constant along the line-of-sight (LOS) within the structure. Furthermore, the optical thickness has a Gaussian dependence on wavelength while the source function is wavelength-independent. The parameters are derived by inverting the contrast profiles of the structure, which are taken in reference to an average absorption profile taken from a nearby quiet-Sun region, free from prominent absorption structures (see e.g., black line in Figure~\ref{fig:ha_prof}a). Typical structures that have been treated with CM are chromospheric mottles or fibrils see \citep[e.g.,][]{tsiropoula97,tziotziou03}, filaments \citep[e.g.,][]{kuckein16}, arch-filament systems \citep{gonzalez_manrique17} and surges \citep{verma20}. In this study, as already mentioned, contrast profiles were processed by applying an iterative PCA process to remove noise before feeding the CM inversion scheme.

To follow the evolution of the minifilament as it expands and erupts, a threshold on the absolute contrast value of the H$\alpha$ profiles over the entire FOV is not sufficient, because other quiet Sun network regions may satisfy this condition and be erroneously counted as part of the minifilament. We set a threshold in the contrast value that generously covered the minifilament and refined it by manually setting a ROI, which contained the minifilament as it expanded, excluding the nearby atmosphere. The drawback of this method is that the investigated region grew with time, but we base it on the assumption that the imprint of the evolving structure within this region will be more easily recognizable, instead of using the entire FOV. For these masks we kept only the good fits and discarded the bad ones \citep[see][for the set of criteria that were used to identify good fits]{dineva20}. 

\begin{figure}
\centering
  \includegraphics[width=9cm]{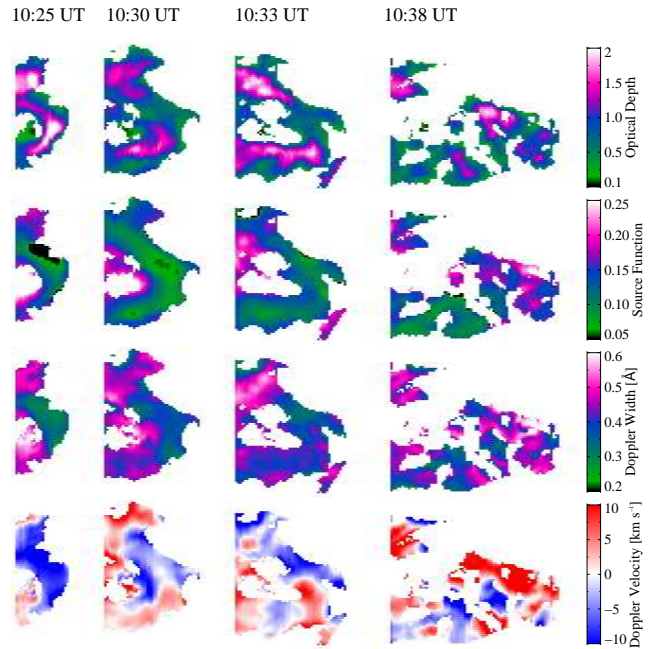}
  \caption
    {Cloud model parameter maps of the minfilament at four instances during its eruption. From top to bottom are the optical depth $\tau$, the source function $S$, the Doppler width $\delta \lambda_{D}$ and the Doppler velocity $\upsilon_{D}$. See text for the mask construction.}
    {\label{fig:cloudmaps}}
\end{figure}

\begin{figure*}
\centering
  \includegraphics[width=18cm]{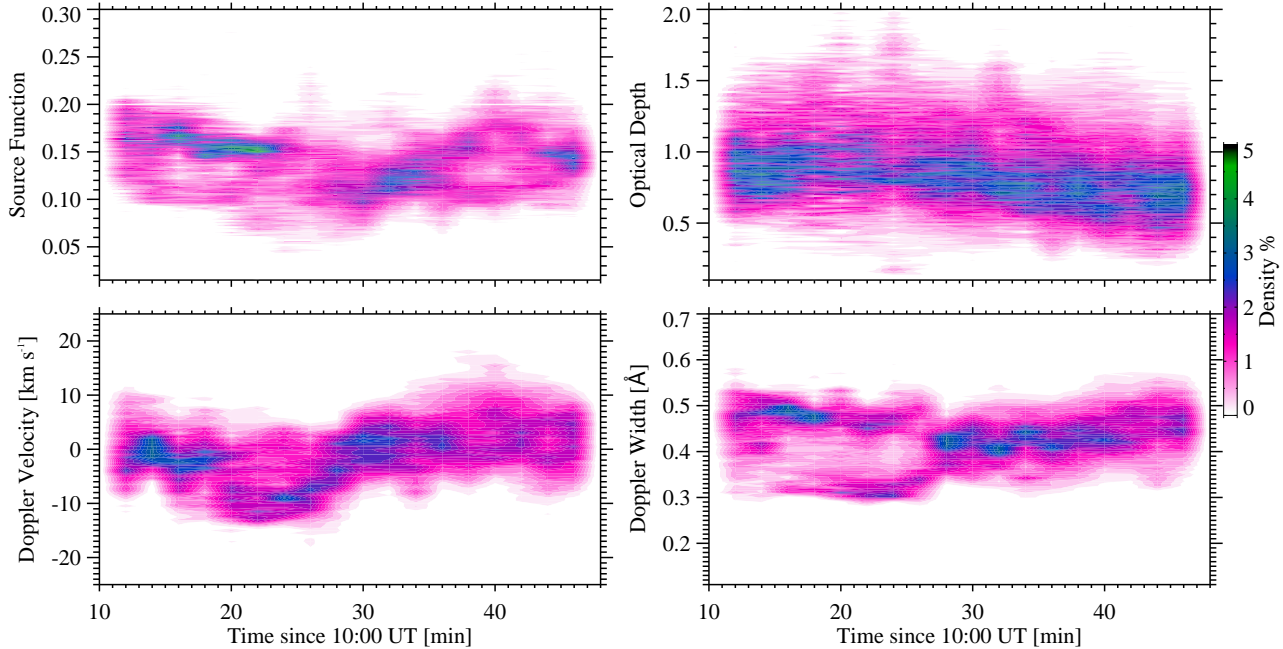}
  \caption
    {Density distribution functions of the CM parameters within the minifilament masks (see text) as a function of time.}
    {\label{fig:cloud2d}}
\end{figure*}

Using these masks, we plot maps of the four CM parameters in Figure~\ref{fig:cloudmaps}. At 10:25\,UT, i.e., just before its eruption, the minifilament has already high optical depth and low source function and Doppler width. Therefore, it consists of optically thick, cool and dense plasma, with the exception of the footpoints and the jet-like feature, which exhibit high Doppler width and are consequently associated with heating. At this instance, the minifilament exhibits a bulk upward motion, with velocities exceeding 10\,km\,s$^{-1}$. The velocities calculated with CM are always higher than the Doppler shifts shown in the maps of Figure~\ref{fig:ha_evo} \citep{chae06,kuckein16,dineva20}.

During the eruption (10:30 -- 10:33\,UT), the optical depth of the minifilament near the apex decreases, as the plasma becomes thinner along the LOS, as a result of the expansion of the structure. However, the arms still maintain their optical thickness. Along the entire minifilament, the source function does not present any notable changes during the eruption. The Doppler width increases in progressively larger parts of the structure, mostly closer to the footpoints, indicating heating of the plasma that constitutes the minifilament. The Doppler velocity maps show that mostly the middle part continues to move upwards, while the footpoints start receding. At 10:33\,UT, the velocity map indicates a more complex flow pattern along the minifilament. Although the middle part maintains the upward motion, the lower part exhibits a pattern of alternating red- and blue-shifts, which persist (10:38\,UT). This pattern is indicative of a three-dimensional motion of the body of the minifilament, possibly exhibiting twisting and writhing as it is stretched due to its eruption. At 10:38\,UT, the shape of this lower part resembles a spiral-like structure, whose decreasing optical depth indicates that it has already started to fade. The Doppler width is higher at the outer parts of the arm than at the middle, which also supports the presence of a complex three-dimensional twisting flow pattern. At the outer parts of the arm, where motions of the spiraling plasma will be roughly along the LOS, velocity gradients will contribute to the Doppler width. In contrast, along the main axis, where these motions are roughly perpendicular to the LOS, this contribution will be much lower. Finally, the regions of the nearby chromosphere ahead of the apex of the minifilament and its upper footpoint are more optically thick, with higher source function, high Doppler width and predominant strong downflows.

Using the minifilament masks constructed for each scan we calculated the two-dimensional density distribution functions of the four CM parameters in the parameter space over time, in two-minute-wide bins (Figure~\ref{fig:cloud2d}). As already mentioned, the area of the masks and, hence, the number of pixels they contained, increased with time, following the expansion of the minifilament. Therefore, we express the density as a percentage of the number of pixels contained in each mask. 

The distribution functions of the CM parameters (Figure~\ref{fig:cloud2d}) show clear evolutionary trends of the erupting minifilament within the density distribution of the background pixels. These trends are visible after 10:20\,UT, when the structure is more clearly seen in the H$\alpha$ scans. As the minifilament appears in the FOV, the source function density distribution shifts to lower values and then increases during the eruption (after 10:20\,UT), before subsiding gradually to the quiet-Sun background of the masks. Conversely, the optical depth exhibits a slight increase but the distribution functions are highly skewed towards high values, as per the appearance of optically thick plasma, reaching values as high as two. These values then tend to decrease during the eruption. The Doppler velocity, initially with no pronounced preferred direction, was predominantly upward after 10:20\,UT, with upward motions in excess of 10\,km\,s$^{-1}$. This upward motion was followed by a bulk downward motion of the minifilament material, which then led to the strong downflows with velocities up to 20\,km\,s$^{-1}$, as discussed also in the previous sections. The Doppler velocities derived with the CM are more representative of the actual motions of the downflowing material. The evolutionary signature of the minifilament is also seen in the density function of the Doppler width. Initially, the minifilament is characterized by narrower absorption profiles than the average quiet-Sun, with Doppler widths equal to $\Delta\lambda =$ 0.3\,{\AA}, with the exception of its footpoints ($\Delta\lambda > 0.5$\,{\AA}). During and after the eruption, spectral profiles with higher Doppler width are more abundant. The profiles associated with the strong downflows are predominantly wider, with Doppler widths exceeding $\Delta\lambda =$ 0.4\,{\AA}.   

It is useful to see how the measured CM parameters of the minifilament compare with other H$\alpha$ absorption structures. For example, \citet{tziotziou03} find that chromospheric mottles have optical depth, source function and Doppler widths around 1.00, 0.15 and 0.45\,{\AA}, respectively, which roughly agree with the ones also reported earlier by \citet{tsiropoula97}. The corresponding average values for arch-filament systems reported by \citet{gonzalez_manrique17} were 1.19, 0.12 and 0.41\,{\AA} while \citet{kuckein16} 
found that the giant filament of their study had corresponding values equal to 1.59, 0.07 and 0.39\,{\AA}. These values, along with the density distributions of Figure~\ref{fig:cloud2d} indicate that minifilaments are closer to quiet-Sun structures when they form but they acquire more filament-like properties during their evolution towards eruption. In that sense, minifilaments stand out from the chromospheric fine-structure contrast background (i.e., fibrils) when they are on their course to eruption, as already noted by \citet{hermans_martin86}.

\section{Discussion and conclusions}\label{sec:conclusions}

A new observing setup at the VTT allowed us to acquire high-spectral resolution scans of extended regions with a fast cadence equal to 20\,s. Using this setup, we were able to capture the eruption of a quiet-Sun minifilament, revealing hitherto unnoticed aspects of their dynamic evolution in the chromosphere and their interaction with neighboring structures.

Spectroscopic inversions based on the CM \citep{beckers64} were performed for the first time for such a structure. The inferred values were close to those of other chromospheric structures but these values started to differ notably when the expansion/eruption process was initiated. Then, the minifilament stood out from the chromospheric ``forest'' and started resembling its large-scale counterparts.

Initially, it contained cool chromospheric plasma and exhibited increasing opacity along with radial expansion. In the chromosphere, we measured a projected speed equal to 2.2\,km\,s$^{-1}$ for the slow expansion and a faster speed of 17.6\,km\,s$^{-1}$ for the eruptive phase. These values are comparable to speeds derived by \citet{panesar20} for a number of jet-like events involving minifilament eruptions. The two phases of eruption, which is also common in active-region filaments, was also supported by the Doppler velocities of the high-resolution H$\alpha$ spectra. During the slow expansion phase, the minifilament was moving upwards as a single structure, before exhibiting a more complicated velocity pattern during the eruption. This pattern comprised alternating blue- and red-shifted patches, which we interpreted as the result of a three-dimensional motion with twisting and writhing components. This intricate velocity structure points to the minifilament being a twisted structure, possibly a small-scale flux rope. The distributions of FWHM and Doppler width, derived via CM inversions, also support this interpretation. However, detailed analyses of more observations in the future can shed more light on the structure and development of flows and instabilities in minifilaments.

During the eruption, a secondary arc protruded from the minifilament, which continued to move upwards before disappearing, implying a thread-like structure of the minifilament. The southern end of this structure interacted with a nearby small-scale structure, causing strong downflows and broadening in the H$\alpha$ profiles. The rest of the minifilament started disappearing, first around its apex and then more towards its footpoints. Part of the material returned to the chromosphere, interacting with nearby small-scale magnetic structures. This was evident by the high values of the BVHM and the FWHM at the wake of the eruption. The corresponding line profiles at these regions were highly asymmetric with broad shoulders in the red wing, indicating strong downflows and velocity gradients, as the minifilament material collided with the low-lying chromospheric structures. Cloud model inversions showed that at these sites downflow velocities exceeded 10\,km\,s$^{-1}$ and Doppler widths surpassed 0.5\,{\AA}. 

The minifilament was associated with flare-like brightenings near its apparent footpoints, another common feature reported in literature \citep[e.g.,][]{ren08,hong11,chen18}. Some of these were seen in EUV, more brightly in 304\,{\AA}, before the start of the slow expansion. Others appeared gradually as the minifilament expanded, before the fast motion. They were also detected in the H$\alpha$ core intensity and they were associated with large FWHM and strong downflows. Such a brightening was also evident at the base of a jet-like structure, which appeared in the northern part of the minifilament. These brightenings can be attributed to reconnection below the minifilament as a result of tether cutting, as the structure stretched the field lines of the constraining magnetic field \citep[see for example the observations of an erupting active region filament in][]{chen18b}. More likely, however, they can be due to the interaction between the minifilament and the small-scale magnetic fields, as the former expanded above them leaving the neutral line region. These reconnection events can act as a driver for the fast expansion that followed \citep{sterling05,panesar20}.        

Regarding the origin of the minifilament, our EUV observations and context H$\alpha$ SJ images suggest that the structure formed during the observations and did not emerge bodily. No magnetic flux emergence of the scale that would justify the appearance of the structure took place at least three hours before the observations. The morphological evolution of the region in EUV suggests a transition from sheared loops to a sigmoid structure. This means that the structure was gradually building up through successive reconnection events. However, lack of H$\alpha$ scans across the entire structure did not allow us to monitor the processes that led to the formation of the minifilament. 

Recent studies attest to the critical role of minifilaments in the dynamics of the quiet Sun and support that common physical mechanisms act both in large-scale and miniature filaments. Similarly to active regions, but at a considerably smaller scale, the magnetic field of the quiet Sun can produce complicated magnetic configurations, amenable to the same instabilities as their active-region counterparts, which can, in turn, give rise to various eruptive phenomena. Findings regarding the abundance of sigmoidal structures in the quiet-Sun and the non-potentiality of the small-scale magnetic fields \citep{chesny13,chesny15,chesny16} support this analogy, albeit coming from a different standpoint. Another pertinent example is the finding that the quiet-Sun magnetic fields of the network obey a fundamental free energy-magnetic helicity relationship \citep{tziotziou14}, similarly to active regions, although in the quiet-Sun magnetic helicity builds mainly because of the action of the shuffling motions on small-scale magnetic fields \citep{tziotziou15}. \citet{chen18} attributed the formation of a minifilament to such motions, resulting to helicity injection after emergence of small-scale magnetic flux. The persistent converging motions of the magnetic footpoints of the region also seem to be involved in the formation and destabilization of the minifilament presented here. This implied analogy between large- and small-scale phenomena may extend to even smaller (perhaps granular) scales \citep[see e.g.,][]{sterling20}, although any conclusive answer will have to wait for regular observations from the new generation of solar telescopes such as the Daniel K. Inouye Solar Telescope \citep[DKIST,][]{Tritschler16} and the European Solar Telescope \citep[EST,][]{jurcak19}. 

A future goal is to capture the evolution of minifilaments during their formation and along their entire length. Further high quality, multi-wavelength observations are needed to decipher their structure, flow fields, and intricate interactions with the chromosphere and corona. To this end, we anticipate the contributions, not only from the new facilities underway, but also from observing setups such as the one used in this study. Such a configuration offers very high spectral resolution combined with fast temporal coverage and good spatial resolution; not only can it provide invaluable insight into the processes that take place in the chromospheric and photospheric environment, but it also offers a more accurate and reliable reference for future magnetohydrodynamics and radiative transfer modelling efforts.

\begin{acknowledgements}
This work was supported by grant DE~787/5-1 of the Deutsche Forschungsgemeinschaft (DFG). ED is grateful for the generous financial support from German Academic Exchange Service (DAAD) in form of a doctoral scholarship. MV and CK acknowledge support by the European Commission's Horizon 2020 Program under grant agreements 824064 (ESCAPE -- European Science Cluster of Astronomy \& Particle physics ESFRI research infrastructures) and 824135 (SOLARNET -- Integrating High Resolution Solar Physics). MV acknowledges the support of DFG grant VE~1112/1-1. The Vacuum Tower Telescope at the Spanish Observatorio del Teide of the Instituto de Astrof\'{\i}sica de Canarias is operated by the German consortium  of the Leibniz-Institut f\"ur Sonnenphysik in Freiburg, the Leibniz-Institut f\"ur Astrophysik Potsdam, and the Max-Planck-Institut f\"ur Sonnensystemforschung G\"ottingen. The authors would like to thank Prof. H. Wang and Dr. A. Warmuth for providing usedul comments on the manuscript.

\end{acknowledgements}

%

\vspace{5mm}
\facilities{VTT, SDO}


\software{SolarSoft IDL \citep{bentley98,Freeland98}, sTools \citep{stools,dineva20}  
      }




\bibliography{references}{}
\bibliographystyle{aasjournal}



\end{document}